%% file: sample-sigconf.tex
\documentclass[sigconf]{acmart}

\usepackage{algorithmic}
\usepackage[ruled, linesnumbered]{algorithm2e}
\newlength\mylen

\usepackage{diagbox}
\usepackage{multirow}
\AtBeginDocument{%
  \providecommand\BibTeX{{%
    \normalfont B\kern-0.5em{\scshape i\kern-0.25em b}\kern-0.8em\TeX}}}
\usepackage{array}
\usepackage{subfigure}
\usepackage{balance}

\setcopyright{acmcopyright}
\copyrightyear{2024}
\acmYear{2024}
\setcopyright{acmlicensed}\acmConference[KDD '24]{Proceedings of the 30th ACM SIGKDD Conference on Knowledge Discovery and Data Mining}{August 25--29, 2024}{Barcelona, Spain}
\acmBooktitle{Proceedings of the 30th ACM SIGKDD Conference on Knowledge Discovery and Data Mining (KDD '24), August 25--29, 2024, Barcelona, Spain}
\acmDOI{10.1145/3637528.3671678}
\acmISBN{979-8-4007-0490-1/24/08}

\begin{document}

\title{Privacy-Preserved Neural Graph  Databases}


\author{Qi Hu}
\affiliation{%
  \institution{Department of CSE, Hong Kong University of Science and Technology}
  \city{Hong Kong}
  \country{China}}
\email{qhuaf@connect.ust.hk}

\author{Haoran Li}
\affiliation{%
  \institution{Department of CSE, Hong Kong University of Science and Technology}
  \city{Hong Kong}
  \country{China}}
\email{hlibt@connect.ust.hk}

\author{Jiaxin Bai}
\affiliation{%
  \institution{Department of CSE, Hong Kong University of Science and Technology}
  \city{Hong Kong}
  \country{China}}
\email{jbai@connect.ust.hk}

\author{Zihao Wang}
\affiliation{%
  \institution{Department of CSE, Hong Kong University of Science and Technology}
  \city{Hong Kong}
  \country{China}}
\email{zwanggc@cse.ust.hk}

\author{Yangqiu Song}
\affiliation{%
  \institution{Department of CSE, Hong Kong University of Science and Technology}
  \city{Hong Kong}
  \country{China}}
\email{yqsong@cse.ust.hk}
\renewcommand{\shortauthors}{Qi Hu, Haoran Li, Jiaxin Bai, Zihao Wang, \& Yangqiu Song}

\input{abstract}





\maketitle

\input{introduction}

\input{related_work}

\input{problem_formulation}

\input{experiments}

\input{conclusion}

\section*{ACKNOWLEDGMENTS}
The authors of this paper were supported by the NSFC Fund (U20B2053) from the NSFC of China, the RIF (R6020-19 and R6021-20) and the GRF (16211520 and 16205322) from RGC of Hong Kong. We also thank the support from UGC Research Matching Grants (RMGS20EG01-D, RMGS20CR11, RMGS20CR12, RMGS20EG19, RMGS20EG21, \\ RMGS23CR05, RMGS23EG08). 

\bibliographystyle{ACM-Reference-Format}
\balance
\bibliography{sample-base}

\input{appendix}

\end{document}

%% file: abstract.tex
\begin{abstract}
In the era of large language models (LLMs), efficient and accurate data retrieval has become increasingly crucial for the use of domain-specific or private data in the retrieval augmented generation (RAG). Neural graph databases (NGDBs) have emerged as a powerful paradigm that combines the strengths of graph databases (GDBs) and neural networks to enable efficient storage, retrieval, and analysis of graph-structured data which can be adaptively trained with LLMs. The usage of neural embedding storage and Complex neural logical Query Answering (CQA) provides NGDBs with generalization ability. When the graph is incomplete, by extracting latent patterns and representations, neural graph databases can fill gaps in the graph structure, revealing hidden relationships and enabling accurate query answering. Nevertheless, this capability comes with inherent trade-offs, as it introduces additional privacy risks to the domain-specific or private databases.  Malicious attackers can infer more sensitive information in the database using well-designed queries such as from the answer sets of where Turing Award winners born before 1950 and after 1940 lived, the living places of Turing Award winner Hinton are probably exposed, although the living places may have been deleted in the training stage due to the privacy concerns. In this work, we propose a privacy-preserved neural graph database (P-NGDB) framework to alleviate the risks of privacy leakage in NGDBs. We introduce adversarial training techniques in the training stage to enforce the NGDBs to generate indistinguishable answers when queried with private information, enhancing the difficulty of inferring sensitive information through combinations of multiple innocuous queries. Extensive experimental results on three datasets show that our framework can effectively protect private information in the graph database while delivering high-quality public answers responses to queries. The code is available at \url{https://github.com/HKUST-KnowComp/PrivateNGDB}.
\end{abstract}

%% file: introduction.tex
\section{Introduction}

Graph DataBases (GDBs) play a crucial role in storing, organizing, and retrieving structured relational data and support many traditional data-intensive applications, such as recommender systems~\cite{dong2018challenges}, fraud detection~\cite{pourhabibi2020fraud}, etc. In the era of Large Language Models (LLMs), the role of GDBs gets increasingly important because of the Retrieval Augmented Generation (RAG) paradigm,\footnote{\url{https://openai.com/research/emergent-tool-use}}\footnote{\url{https://docs.llamaindex.ai/en/stable/examples/agent/openai_agent.html}}\footnote{\url{https://www.langchain.com/use-case/agents}}  where the LLM agents can be significantly enhanced by external GDBs such as Knowledge Graphs (KGs)~\cite{edge2024local, mavromatis2024gnn,gao2023retrieval}. Such interaction of LLMs and GDBs enables new possibilities by creating interactive natural language interfaces over structural data for domain-specific applications.


The RAG paradigm of LLMs also highlights new challenges of traditional GDB. The natural language interaction enlarges the possible ways of utilizing the knowledge managed in GDBs, requiring simultaneous knowledge discovery and query answering. Knowledge discovery is particularly important for domain-specific data and knowledge graphs, as the graph database may not capture all the necessary relationships and connections between entities by traversing~\cite{bordes2013translating, galarraga2013amie}. 

\begin{figure*}[ht]
  \centering
  \includegraphics[width=0.99\textwidth]{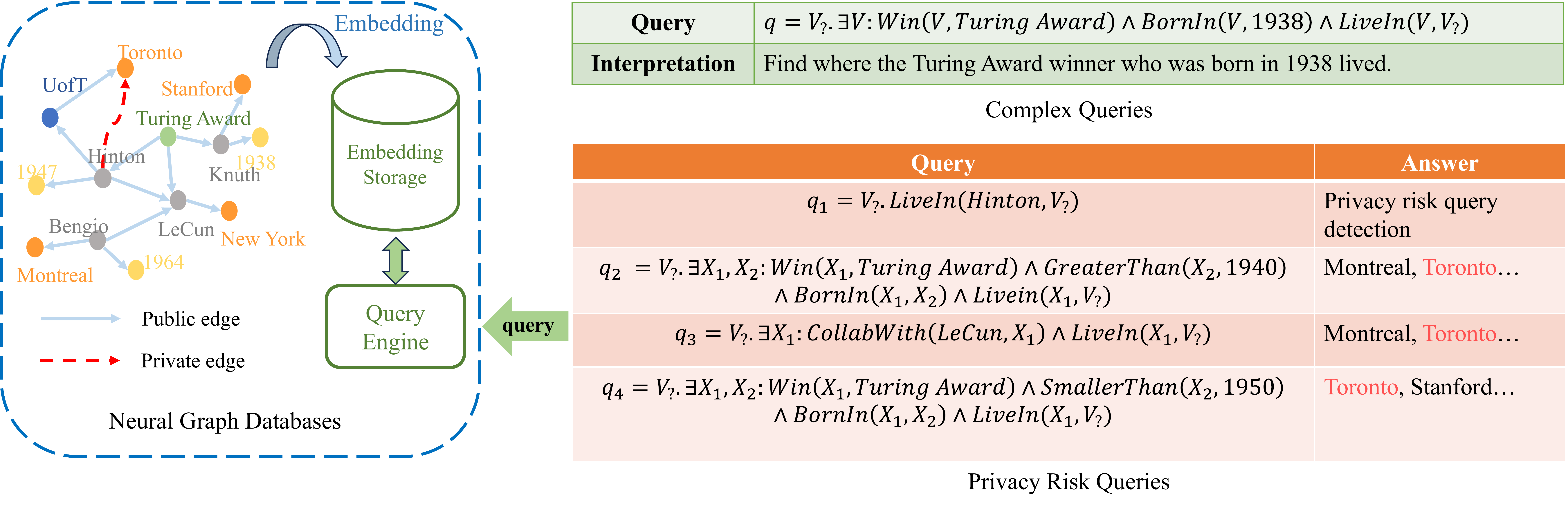}
  \caption{Privacy risks of NGDBs facing malicious queries. 
  To illustrate the issue, consider an example where an attacker attempts to infer private information about Hinton's living place in the NGDBs. Direct querying private information can be easily detected by privacy risk detection, however, attackers can leverage well-designed queries to retrieve desired privacy. In this example, the privacy attacker can query ``where Turing Award winners born after 1940 lived,'' ``where ones before 1950 lived,'' and ``where LeCun's collaborator lived,'' etc. And the query can be more complex as shown in privacy risk queries. The return answer denoted in red may leak private information in  NGDBs. The intersection of these queries still presents a significant likelihood of exposing the living place of Turing Award winner Hinton.}
  \label{Fig:intro}
\end{figure*}

NGDBs have been proposed to address these challenges by extending the concept of graph DBs~\cite{ren2023neural,besta2022neural}. They combine the flexibility of graph data models with the computational capabilities of neural networks, enabling effective and efficient representation, storage, and analysis of interconnected data. 
As shown in Figure~\ref{Fig:intro}, NGDBs are designed with two modules: neural graph storage providing unified storage for diverse entries in an embedding space and neural query engine searching answers to input complex queries from the unified storage~\cite{ren2023neural}. By utilizing graph-based neural network techniques, these databases empower expressive and intelligent querying, inference, and knowledge discovery. This capability, known as complex query answering (CQA)~\cite{arakelyan2020complex}, aims to identify answers that satisfy given logical expressions~\cite{hamilton2018embedding}. 
Query encoding methods are commonly used in CQA. These methods involve parameterizing entities, relations, and logical operators, encoding both queries and entities into the same embedding space simultaneously, and retrieving answers according to the similarity between the query (or set) embeddings and answer embeddings. For example, GQE~\cite{hamilton2018embedding},  Q2B~\cite{ren2020query2box}, and Q2P~\cite{bai2022query2particles} encode queries as vectors, a hyper-rectangle, and a set of particles, respectively. 
Ideally, CQA can be jointly trained with RAG so that the language model agents can better adapt to domain-specific data~\cite{DBLP:conf/acl/LeeCT19,DBLP:conf/icml/GuuLTPC20}.

While NGDBs and their CQA ability have demonstrated significant achievements in various domains, they do face unique privacy challenges in comparison to traditional graph DBs~\cite{zeighami2023neurodb,zeighami2023neural}. One notable risk arises from their generalization ability. Although generalization efficiently handles incomplete knowledge and enriches the retrieved information, it also enables attackers to infer sensitive information from NGDBs by leveraging the composition of multiple complex queries. Figure~\ref{Fig:intro} demonstrates an example of privacy attacks via malicious complex queries: the living place information of an individual. In this case, {\it Hinton} as an example, is considered to be private. Even though Hinton's residence is omitted during the construction of the knowledge graph, or if direct privacy queries are restricted, a malicious attacker can still infer this sensitive information without directly querying Hinton's living place but conducting well-designed related queries. The answers denoted in red may leak sensitive information by accident.

Graph privacy has consistently been a significant concern in the field of graph research. Multiple studies demonstrate the vulnerability of graph neural networks to various privacy leakage issues. For instance, graph embeddings are vulnerable to attribute inference attacks~\cite{duddu2020quantifying, gong2018attribute, zhang2022inference} and link prediction attacks~\cite{he2021stealing, wu2022linkteller}, and various protection methods are proposed~\cite{hu2023independent, hu2022learning, wang2021privacy}. However, these works focus on graph representation learning and have the assumption that the graph embeddings can be reached by the attackers which are not typical scenarios in NGDBs. NGDBs often provide CQA service and do not publish the learned embeddings, and attackers can only infer private information using thoughtfully designed queries. Furthermore, these works solely study node classification and link prediction problems, leaving the query answering privacy in NGDBs unexplored, resulting in the potential privacy leakage problem in LLMs~\cite{li2023privacy}.

In this paper, we expose the potential privacy risks of NGDBs with formal definition and evaluation. We introduce privacy protection objectives to the existing NGDBs, categorizing the answers of knowledge graph complex queries into private and public domains. To safeguard private information, NGDBs should strive to maintain high-quality retrieval of non-private answers for queries, while intentionally obfuscating the private-threatening answers to innocuous yet private queries. As shown in Figure~\ref{Fig:intro}, the answer sets of queries proposed by privacy attackers are obfuscated, and attackers will face greater challenges in predicting private information, thereby effectively safeguarding the privacy of NGDBs. Meanwhile, we also create the corresponding benchmark on three datasets (Freebase~\cite{bollacker2008freebase}, YAGO~\cite{suchanek2007yago}, and DBpedia~\cite{bizer2009dbpedia}) to evaluate the performance of CQA on public query answers and the protection of private information. 

To alleviate the privacy leakage problem in NGDBs, we propose Privacy-preserved Neural Graph Databases (P-NGDBs) as a solution. P-NGDBs divide the information in graph databases into private and public parts and categorize complex queries' answers to various privacy risks based on whether or not they involve sensitive information. P-NGDBs can provide answers with different levels of precision in response to queries with varying privacy risks. We introduce adversarial techniques in the training stage of P-NGDBs to generate indistinguishable answers when queried with private information, enhancing the difficulty of inferring privacy through complex private queries. We summarize our major contributions as follows:
\begin{itemize}
    \item To the best of our knowledge, our work represents the pioneering effort in investigating privacy leakage issues of complex queries in NGDBs and providing formal definitions for privacy protection in this domain.
    \item Based on three public datasets, we propose a benchmark for the simultaneous evaluation of CQA ability and privacy protection for NGDBs.
    \item We introduce P-NGDB, a privacy protection framework for safeguarding privacy in NGDBs. Extensive experiments conducted on three datasets demonstrate its effectiveness in preserving retrieval performance and protecting privacy.
\end{itemize}

The rest of the paper is organized as follows. Section \ref{sec:related} systematically reviews the related work. Section \ref{sec:preliminary} introduces the preliminary and definition of the privacy problem in NGDBs. Section \ref{sec:benchmark} and \ref{sec:problem_sec} introduce the innocuous yet private queries and the framework of P-NGDBs in detail, respectively. Section \ref{sec:exp} evaluates the performance of P-NGDBs on the benchmark based on three real-world datasets. Finally, we conclude our work in Section \ref{sec:conclude}.  

%% file: related_work.tex
\section{Related Work \label{sec:related}}

\subsection{Complex Query Answering}
One essential function of neural graph databases is to answer complex structured logical queries according to the data stored in the NGDBs, namely complex query answering (CQA)~\cite{wang2022logicalsurvey,ren2023neural}. Query encoding is one of the prevalent methods for complex query answering because of its effectiveness and efficiency.
These methods for query embedding utilize diverse structures to represent complex logical queries and effectively model queries across various scopes. Some methods encode queries as various geometric structures: such as vectors~\cite{hamilton2018embedding,bai2022query2particles}, boxes~\cite{ren2020query2box,liu2021neural}, cones~\cite{zhang2021cone} and cones in hyperbolic spaces~\cite{liu2023poine}. Other methods leverage the probability distributions such as Beta~\cite{ren2020beta}, Gaussian~\cite{choudhary2021probabilistic}, and Gamma~\cite{yang2022gammae} distributions to encode logic knowledge graph queries. Meanwhile, some approaches motivate the embedding space from the fuzzy logic theory~\cite{chen2022fuzzy,wang2023wasserstein}.
Advanced neural structures are also utilized to encode complex queries: such as transformers and sequential encoders~\cite{kotnis2021answering,bai2023sequential}, graph neural networks~\cite{zhu2022neural}. Pretrained shallow neural link predictors are also adapted into search algorithms~\cite{arakelyan2020complex,bai2023answering,yin2024rethinking}, and learning-to-search frameworks~\cite{wang2022logical,xu2023query2triple}.
Notably, NRN~\cite{bai2023knowledge} is proposed to handle CQA with numerical values, which is of particular interest in privacy protection.

While complex query answering has been widely studied in past works, the privacy problem in NGDBs is overlooked. With the development of the representation ability in NGDBs, the issue of privacy leakage has become increasingly critical. There are some works showing that graph representation learning is vulnerable to various privacy attacks~\cite{hu2022learning}. \cite{salem2018ml, olatunji2021membership} show that membership inference attacks can be applied to identify the existence of training examples. Model extraction attacks~\cite{tramer2016stealing} and link stealing attacks~\cite{he2021stealing, zhang2021link} try to infer information about the graph representation model and original graph link, respectively. However, these works assume that attackers have complete access to the graph representation models. In NGDBs, however, sensitive information can be leaked during the query stage, presenting a different privacy challenge. Our research explores the issue of privacy leakage in NGDBs and presents a benchmark for comprehensive evaluation.

\subsection{Graph Privacy}

Various privacy protection methods have been proposed to preserve privacy in graph representation models. Anonymization techniques~\cite{liu2008towards, rossi2015k, hoang2020cluster, zheleva2007preserving} are applied in graphs to reduce the probability of individual and link privacy leakage. Graph summarization~\cite{hay2008resisting} aims to publish a set of anonymous graphs with private information removed. Learning-based methods are proposed with the development of graph representation learning, \cite{wang2021privacy, liao2020graph,li2020adversarial} regard graph representation learning as the combination of two sub-tasks: primary objective learning and privacy preservation, and use adversarial training to remove the sensitive information while maintaining performance in the original tasks. Meanwhile, some methods~\cite{liu2022fair, oh2022learning, hu2023independent} disentangle the sensitive information from the primary learning objectives. Additionally, differential privacy~\cite{kasiviswanathan2013analyzing, day2016publishing, shen2013mining} introduces noise into representation models and provides privacy guarantees for the individual privacy of the datasets~\cite{daigavane2021node}. Though noise introduced by differential privacy protects models from privacy leakage, it can also impact the performance of the original tasks significantly. To empirically evaluate the privacy leakage problem in representations, some works are proposed to construct benchmarks~\cite{li2023p}. Federated learning utilizes differential privacy and encryption to prevent the transmission of participants' raw data and is widely applied in distributed graph representation learning~\cite{peng2021differentially, he2021fedgraphnn,hu2023user}. However, federated learning cannot be applied to protect intrinsic private information in the representation models.

While there are various graph privacy preservation methods, they only focus on simple node-level or edge-level privacy protection. However, the privacy risks associated with neural graph databases during complex query answering tasks have not received sufficient research attention. The development of CQA models introduces additional risks of privacy leakage in NGDBs that attackers can infer sensitive information with multiple compositional queries. Our proposed P-NGDB can effectively reduce the privacy leakage risks of malicious queries.

%% file: problem_formulation.tex
\section{Preliminary and Problem Formulation \label{sec:preliminary}}

\subsection{Preliminary}
We denote a knowledge graph as $\mathcal{G} = (\mathcal{V}, \mathcal{R}, \mathcal{A})$, where $\mathcal{V}$ denotes the set of vertices representing entities and attribute values in the knowledge graph.
$\mathcal{R}$ denotes the set of relations in the knowledge graph. $\mathcal{A}$ denotes the set of attributes that can be split into private attributes and public attributes: $\mathcal{A} = \mathcal{A}_{\text{private}} \cup \mathcal{A}_{\text{public}}$, where $a(u,x) \in \mathcal{A}$ is a triplet where $u,x \in \mathcal{V}$, describing the attribute of entities. $a(u,x)$ denotes that the entity $u$ has the attribute $a$ of value $x \in \mathcal{V}$. 
To prevent privacy leakage, the private attribute of entities, we denote the  $\mathcal{A}_{\text{private}}$ cannot be exposed and should be protected from inferences.

The complex query is the key task of NGDBs and can be defined in existential positive first-order logic form, consisting of various types of logic expressions like existential quantifiers $\exists$, logic conjunctions $\wedge$, and disjunctions $\vee$. In the logical expression, there is a unique variable $V_?$ in each logic query that denotes our query target. Variables $V_1, \cdots, V_k$ denote existentially quantified entities in a complex query. Besides, there are anchor entities $V_a$ and values $X_a$ with the given content in a query. The complex query aims to identify the target entity $V_?$, such that there are $V_1, \cdots, V_k \in \mathcal{V}$ in the knowledge graph that can satisfy the given logical expressions. We denote complex query expression in the disjunctive normal form (DNF) in the following:

\begin{equation}
\begin{split}
    q[V_?] &= V_?. V_1,...,V_k:c_1 \lor c_2 \lor ... \lor c_n \\
c_i &= e_{i,1} \land e_{i,2} \land ... \land e_{i,m},
\end{split}
\end{equation}
where $e_{i,j}$ is the atomic logic expression, which can be $r(V, V')$ denotes relation $r$ between entities $V$ and $V'$, and $a(V, V')$ denotes attribute $a$ of entities $V$ with value $V'$. $c_i$ is the conjunction of several atomic logic expressions $e_{i,j}$. $V, V'$ are either anchor entities or attributes, or existentially quantified variables.

\subsection{Problem Formulation}
Suppose that there is a graph $\mathcal{G} = (\mathcal{V}, \mathcal{R}, \mathcal{A})$, where part of entities' attributes are regarded as private information $\mathcal{A} = \mathcal{A}_{\text{private}} \cup \mathcal{A}_{\text{public}}$, where $a_p(u,x) \in \mathcal{A}_\text{private}$ denotes that the entity $u$ has the sensitive attribute $a_p$ with value $x$. Due to the variation in privacy requirements, an attribute can be handled differently for various entities. NGDBs store the graph in embedding space and can be queried with arbitrary complex logical queries. Due to the generalization ability of NGDBs, attackers can easily infer sensitive attributes $\mathcal{A}_{\text{private}}$ utilizing complex queries. To protect private information, NGDBs should retrieve obfuscated answers when queried with privacy risks while preserving high accuracy in querying public information.

\section{Innocuous yet Private Query \label{sec:benchmark}}
As there is sensitive information in the neural graph databases, some specified complex queries can be utilized by attackers to infer privacy, which is defined as Innocuous yet Private (IP) queries. Some answers to these queries can only be inferred under the involvement of private information, which is a risk to privacy and should not be retrieved by the NGDBs.

\begin{figure}[tbp]
  \centering
  \includegraphics[width=0.99\linewidth]{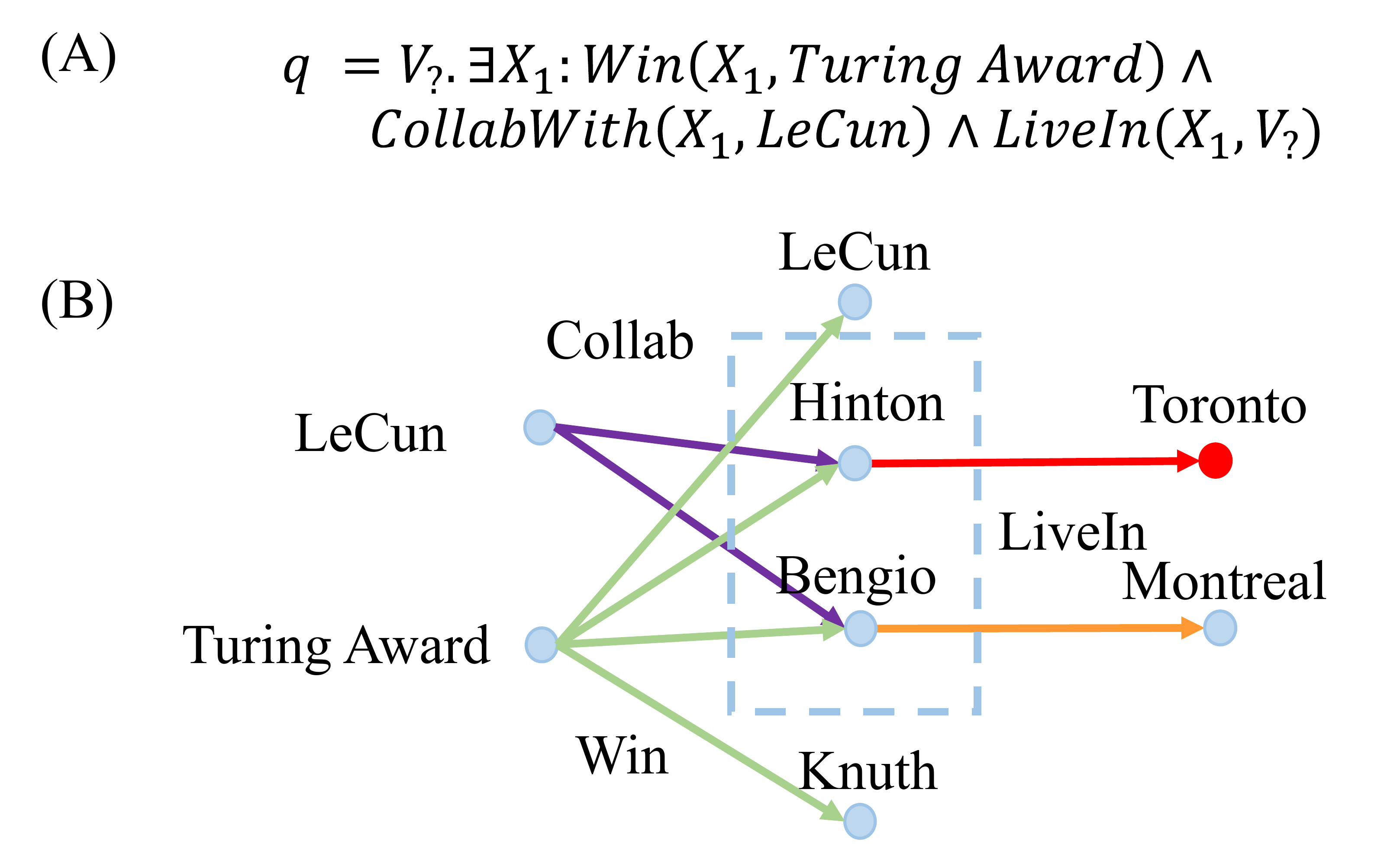}
  \caption{An example of query demonstrating the retrieved privacy-threatening query answers. The orange node denotes privacy risks. (A) The logic knowledge graph query involves privacy information. (B) An example of a complex query in the knowledge graph. Toronto is regarded as a privacy-threatening answer as it has to be inferred by sensitive information.
  %
  }
  \label{fig:illustration}
\end{figure}
\begin{figure*}[ht]
  \centering
  \includegraphics[width=0.85\textwidth]{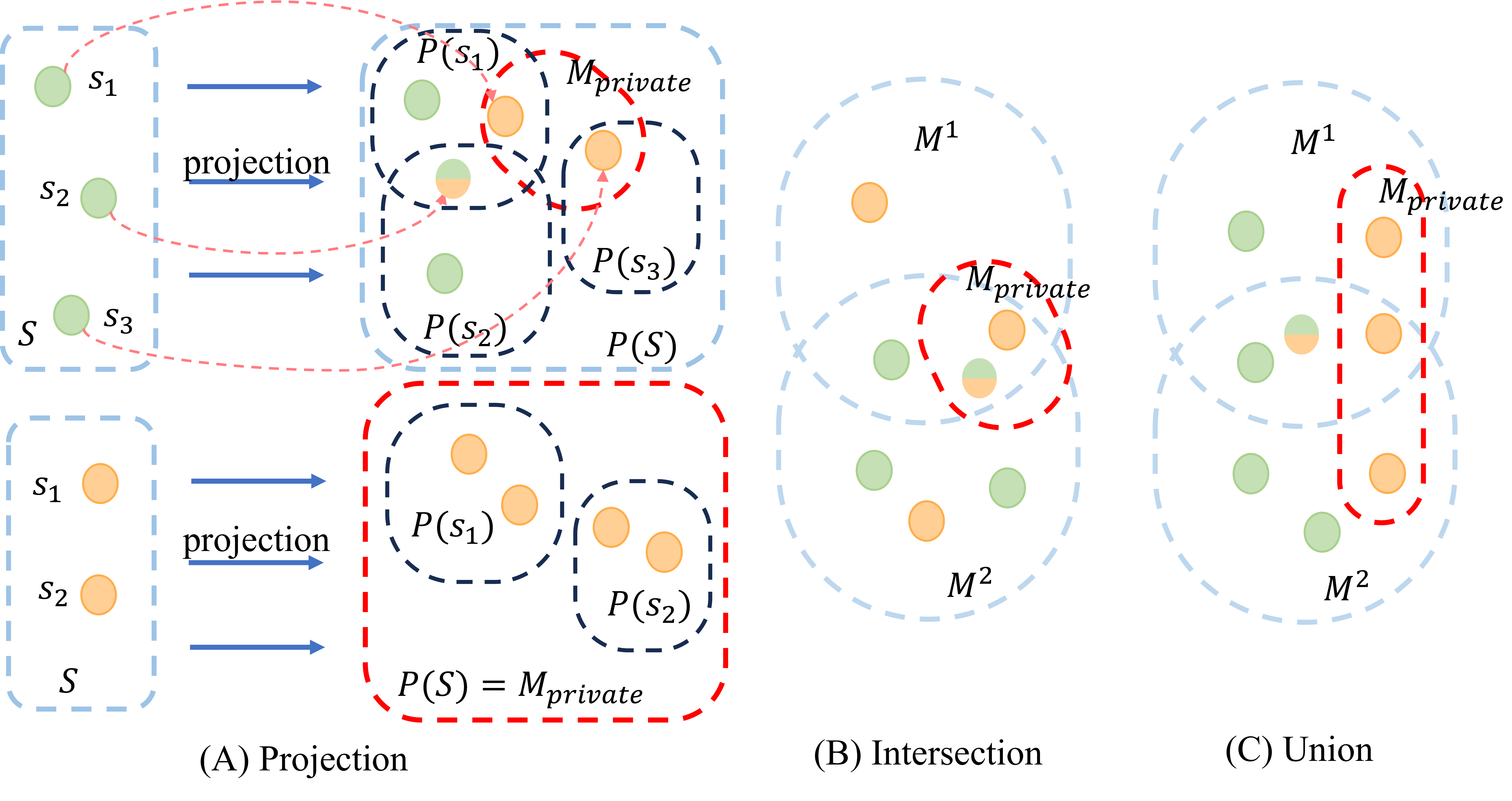}
  \caption{Example of privacy-threatening answer sets computation in projection, intersection, and union. Green nodes denote non-private answers, orange nodes denote privacy-threatening answers, and green-orange nodes denote different privacy risks in subsets. Red dashed arrows denote privacy projection. The range demarcated by the red dashed lines denotes privacy-threatening answer sets. The answers circled in red dashed line are at risk of leaking privacy.}
  \label{Fig:operator}
\end{figure*}

\subsection{Computational Graph}
The query $q$ can be parsed into a computational graph as a directed acyclic graph, which consists of various types of directed edges that represent operators over sets of entities and attributes.
\begin{itemize}
    \item \textit{Projection}: Projection has various types for attribute- and value-sensitive queries~\cite{bai2023knowledge}. Given a set of entities $S$, a relation between entities $r \in \mathcal{R}$, an attribute type $a$, a set of values $S \subset \mathcal{V}$, projection describes the entities or values can be achieved under specified relation and attribute types. For example, attribute projection is denoted as $P_a(S)=\{x \in  \mathcal{V} | v \in S, a(v,x) \in \mathcal{A}\}$ describing the values that can be achieved under the attribute type $a$ by any of the entities in $S$;
    Furthermore, we have \textit{relational projection} $P_r (S)$, \textit{reverse attribute projection} $P_a ^{-1}(S)$. We denote all these projections as $P(S)$;
    \item \textit{Intersection}: Given a set of entity or value sets $S_1,S_2,..., S_n$, this logic operator  computes the intersection of those sets as  $ \cap_{i=1}^n S_i  $;
    \item \textit{Union}: Given a set of entity or value sets $S_1  ,S_2  ,...,S_n$, the operator computes the union of $ \cup_{i=1}^n S_i$. 
\end{itemize}

\subsection{Privacy Threatening Query Answers\label{sec:risk}}
After a complex query is parsed into the computational graph $G$, and $M=G(S)$ denotes the answer set of entities or values retrieved by the computational graph. We divide the $M$ retrieved by the complex queries as private answer set $M_{\text{private}}$ and non-private answer set $M_{\text{public}}$ based on whether the answer has to be inferred under the involvement of private information.  As shown in Figure~\ref{fig:illustration}, query in (A) retrieves answers from the knowledge graph, Toronto is regarded as a privacy-threatening answer because the answer {\it Toronto} denotes the private information $LiveIn(Hinton, Toronto)$. 
To analyze the risks, we define private answer sets for different operators as follows. We denote the input set of operators as $S = S_{\text{private}} \cup S_{\text{public}}$, where $S_{\text{private}}$, $S_{\text{public}}$ denote whether the elements in the set involve private information. 

\textbf{\textit{Projection}}: A projection operator's output will be regarded as private if the operation is applied to infer private attributes. Assume that $M = P(S)$ and input $S = S_{\text{private}} \cup S_{\text{public}}$. $P(S) = P(S_{\text{private}})\cup P(S_{\text{public}})$, we use $p$ denote the corresponding attribute of projection $P$. Without loss of generalization, we discuss the formal definition of projection private answer set $M_{\text{private}}$ in two scenarios. For the projection on public inputs:
\begin{align*}
M_{\text{private}} = \{m \mid \exists v \in S: m \in M \land p(v,m) \in \mathcal{A}_{\text{private}} \land \\
p(u,m) \neq p(v,m), \forall u \in S/\{v\}, p(u,m) \notin \mathcal{A}_{\text{private}}\}.
\end{align*}
If an answer can only be inferred from private information, it will be regarded as a privacy threatening answer.
For the projection on private inputs, 
$$
M_{\text{private}} = P(S_{\text{private}}).
$$
The answers of projection on private inputs are privacy threatening answers as they can expose the existence of private inputs. As shown in Figure~\ref{Fig:operator}(A), all the orange nodes within the answer sets can be solely deduced from private attribute links and are classified as query answers posing privacy risks. On the other hand, nodes such as the green-orange node, while accessible through private projection, can also be inferred from public components, thereby carrying lower privacy risks.

\textbf{\textit{Intersection}}: An intersection operator's output will be regarded as private if the answer belongs to any of the private answer sets. Given a set of answer sets $M^1, M^2, ..., M^n$, where each answer set $M^i = M^i_{\text{private}}\cup M^i_{\text{public}}$, after intersection operator, $M_{\text{private}} = \cap_{i=1}^n M^i - \cap_{i=1}^n M^i_{\text{public}} $. As shown in Figure~\ref{Fig:operator}(B), the green-orange node is categorized as a privacy-threatening answer because it denotes the existence in the answer subsets.

\textbf{\textit{Union}}: A union operator's output will be regarded as private if the answer is the element that belongs to the private answer set of computational subgraphs while not belonging to public answer sets. Given a set of answer sets $M^1, M^2, ...,M^n$, $M_{\text{private}} = \cup_{i=1}^n M^i_{\text{private}} - \cup_{i=1}^n M^i_{\text{public}} $. As shown in Figure~\ref{Fig:operator}(C), all the orange nodes within the answer sets can only be inferred from private attribute links and are classified as privacy-threatening answers.

\section{Privacy-Preserved Neural Graph  Database\label{sec:problem_sec}}
In this section, we mainly focus on the query encoding methods in NGDBs and present the privacy-preserved neural graph databases for protecting sensitive information in knowledge graphs while preserving high-quality complex query answering performance. We set two optimization goals for P-NGDBs: preserve the accuracy of retrieving non-private answers and obfuscate the privacy-threatening answers. 

\subsection{Encoding Representation}
The query encoding methods can be decomposed into two steps: encode queries to embedding space $q \in \mathbb{R}^d$ and compare the similarity with the entities or attributes for answer retrieval. 
Given a query, we iteratively compute the queries based on the sub-query embeddings and logical operators. Assume the encoding process in step $i$, the sub-query embedding is $q_i$. We can denote the logical operators projection, intersection, and union as:
\begin{align}
    q_{i+1} &= f_P(q_i, r), \quad r \in \mathcal{R}\cup \mathcal{A}, \\
    q_{i+1} &= f_I(q^1_i,...,q^n_i),\\
    q_{i+1} &= f_U(q^1_i,...,q^n_i),
\end{align}
where $f_P$, $f_I$, and $f_U$ denote the parameterized projection, intersection, and union operators, respectively. Here, the $f_P$, $f_I$, and $f_U$ can be instantiated with various types of parameterization of query encoding functions. We list all the instantiations used in our experiments in the Appendix~\ref{apdx:qe_instantiations}. After query encoding, we compute the score for every candidate based on the query encoding and entity (attribute) embeddings. Finally, we calculate the normalized probability using the Softmax function:
\begin{equation}
    p(q, v) = e^{s(q,e_v)} / \sum_{u \in \mathcal{C}}e^{s(q,e_u)},
\end{equation}
where $s$ is the scoring function, like similarity, distance, etc., and $\mathcal{C}$ is the candidate set.

\subsection{Learning Objective}
In privacy-preserved NGDBs, there are two learning objectives: given a query, P-NGDBs should accurately retrieve non-private answers and obfuscate the private answers. For public retrieval, given query embedding $q$, the loss function can be expressed as:
\begin{equation}
    L_u =-\frac{1}{N} \sum_{v \in \mathcal{M}_\text{public}^q} \log p(q, v),
\end{equation}
where $\mathcal{M}_\text{public}^q$ is the public answer set for the given query.

While for privacy protection, the learning objective is antithetical. Instead of retrieving correct answers, the objective is to provide obfuscated answers to address privacy concerns. Therefore for private answers, given query embedding $q$, the objective can be expressed as:
\begin{equation}
    \hat{\theta}_g = \arg\min_{\theta_g} \sum_{v \in \mathcal{M}_\text{private}^q} \log p(q,v),
\end{equation}
where $\mathcal{M}_\text{private}^q$ is the private risk answer set for given query.

As discussed in Section \ref{sec:risk}, the query can be decomposed into sub-queries and logical operators. For intersection and union, the logical operators do not generate new privacy invasive answers, while for projection, the answer can be private if the projection is involvement of private information. Therefore, we can directly optimize the projection operators to safeguard sensitive information. The privacy protection learning objective can be expressed as:
\begin{equation}
    L_p = \frac{1}{|\mathcal{A}_\text{private}|} \sum_{r(u,v) \in \mathcal{A}_\text{private}} \log p(f_p(e_v, r), u).
\end{equation}
We aim to reach the two objectives simultaneously and the final learning objective function can be expressed as:
\begin{equation}
    L = L_u + \beta L_p,
\end{equation}
where $\beta$ is the privacy coefficient controlling the protection strength in P-NGDBs, larger $\beta$ denotes stronger protection.

%% file: experiments.tex
\section{Experiments \label{sec:exp}}
In this section, to evaluate the privacy leakage problem in NGDBs and the protection ability of our proposed Privacy-preserved Neural Graph Database (P-NGDBs), we construct a benchmark on three real-world datasets and evaluate P-NGDB's performance based on that.

\begin{figure}[t]
  \centering
  \includegraphics[width=0.95\linewidth]{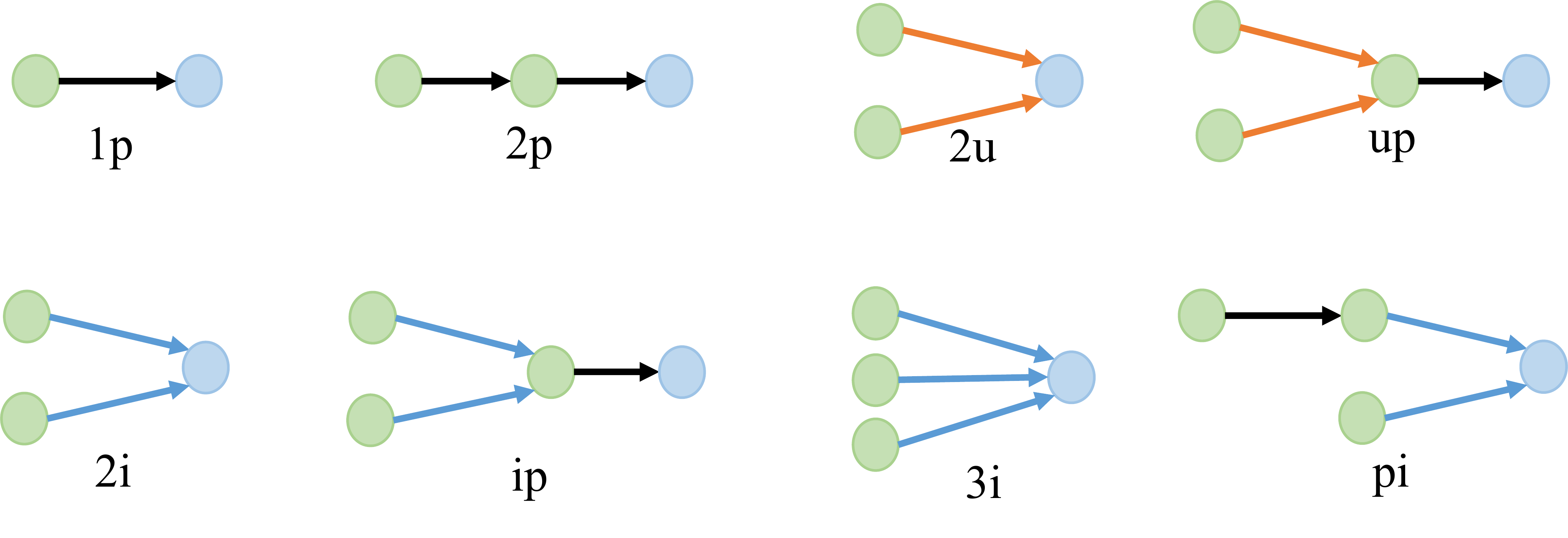}
    \vspace{-0.2cm}
  \caption{Eight general query types. Black, blue, and orange arrows denote projection, intersection, and union operators respectively.}
  \label{fig:query_types}
  \vspace{-0.4cm}
\end{figure}

\subsection{Datasets}
There are various knowledge graph datasets used in complex query answering tasks~\cite{bollacker2008freebase, carlson2010toward}. Without loss of generality, we consider the privacy problem in the three commonly used multi-relational knowledge graph with numerical attributes: FB15K-N, DB15K-N, and YAGO15K-N~\cite{kotnis2018learning} for several reasons: First, attribute value projections can be the same as traditional relation projection if the values themselves are entities, e.g., locations; Second, attributes and their values are more aligned with real-world privacy considerations; Third, attribute values are vulnerable to be attacked as we can use group queries to attack individual's information, which has been widely used as an illustration in differential privacy~\cite{dwork2008differential}. 
We create our benchmarks based on the numerical complex query datasets proposed in~\cite{bai2023knowledge}. In each knowledge graph, there are vertices describing entities and attributes and edges describing entity relations, entity attributes, and numerical relations. We first randomly select a set of edges denoting entities' attributes as privacy information. Then We divide the remaining edges with a ratio of 8:1:1 to construct training, validation, and testing edge sets, respectively. Then the training graph $\mathcal{G}_{train}$, validation graph $\mathcal{G}_{val}$, and testing graph $\mathcal{G}_{test}$ are constructed on training edges, training+validation edges and training+validation+testing edges, respectively. The detailed statistics of the three knowledge graphs are shown in Table \ref{tab:kg_stats}, \#Nodes represents the number of entities and attributes. \#Edeges represents the number of relation triples and attribute triples. \#Pri. Edges represent the number of attribute triples that are considered private. 

\begin{table}[t]
\caption{The statistics of the three knowledge graphs and sampled private attributes.}
\label{tab:kg_stats}
\begin{tabular}{ccccc}
\toprule
Graphs & Data Split & \#Nodes &  \#Edges & \#Pri. Edges \\
\midrule
\multirow{3}{*}{FB15k-N} & Training & 22,964 & 1,037,480 & \multirow{3}{*}{8,000}  \\
 & Validation & 24,021 & 1,087,296 &  \\
 & Testing & 27,144 & 1,144,506 &  \\
 \midrule
\multirow{3}{*}{DB15k-N} & Training & 27,639 & 340,936 & \multirow{3}{*}{6,000} \\
 & Validation & 29,859 & 381,090 &  \\
 & Testing & 36,358 & 452,348 & \\
 \midrule
\multirow{3}{*}{YAGO15k-N} & Training & 30,351 & 383,772 &\multirow{3}{*}{1,600} \\
 & Validation & 31,543 & 417,356 & \\
 & Testing & 33,610 & 453,688 &  \\
\bottomrule
\end{tabular}
\end{table}

\begin{table*}[t]
\caption{The statistics of the testing queries' answers sampled from the three knowledge graphs.}
\label{tab:queries_stats}
\begin{tabular}{ccccccccccc}
\toprule
Graphs & \#Test Ans. & 1p & 2p & 2i & 3i & pi & ip & 2u & up & All \\
\midrule
\multirow{2}{*}{FB15k-N} & Public & 42,952 & 768,012 & 138,446 & 68,677 & 1,120,065 & 307,546 & 1,247,120 & 1,844,682 & 5,537,500 \\
 & Private & 524 & 65,212 & 1,637 & 2,576 & 322,928 & 30,927 & 779 & 239,176 & 663,099 \\
 \midrule
\multirow{2}{*}{DB15k-N} & Public & 16,776 & 764,899 & 26,836 & 20,426 & 1,050,883 & 105,096 & 1,296,453 & 1,959,063 & 5,240,432 \\
 & Private & 211 & 25,236 & 677 & 763 & 148,769 & 6,067 & 266 & 146,218 & 328,207 \\
 \midrule
\multirow{2}{*}{YAGO15k-N} & Public & 14,880 & 913,082 & 38,546 & 23,066 & 1,553,590 & 164,430 & 1,111,663 & 1,920,157 & 5,739,414 \\
 & Private & 136 & 2,319 & 499 & 1742 & 3,656 & 1,483 & 175 & 40,267 & 50,277 \\
 
 \bottomrule
\end{tabular}
\end{table*}

\subsection{Benchmark Construction}
Following previous work~\cite{hamilton2018embedding, bai2023knowledge, wang2021benchmarking}, we evaluate the complex query answering performance on the following eight general query types with abbreviations of $1p,2p,2i,3i,ip,pi,2u$, and $up$, as shown in Figure~\ref{fig:query_types}. For each general query type, each edge represents either a projection or a logical operator, and each node represents either a set of entities or numerical values. We use the sampling method proposed in~\cite{bai2023knowledge} to randomly sample complex queries from knowledge graphs. We randomly sample training, validation, and testing queries from the formerly constructed graphs respectively. For the training queries, we search for corresponding training answers on the training graph. For the validation queries, we use those queries that have different answers on the validation graph from answers on the training graph to evaluate the generalization ability. For the testing queries, we conduct a graph search on the testing graph with private edges for testing answers and we split the answers into private set and public set according to the definition of privacy risk query answers discussed in Section \ref{sec:benchmark}. We conduct a statistical analysis of the number of privacy risk answers for different types of complex queries of three knowledge graphs and the statistics are shown in Table \ref{tab:queries_stats}.

\subsection{Experimental Setup}

\subsubsection{Baselines} Our proposed P-NGDBs can be applied to various complex query encoding methods to provide privacy protections. We select three commonly used complex query encoding methods and compare the performance with and without P-NGDB's protection. To better encode the numerical attribute values, we also use  NRN~\cite{bai2023knowledge}  in the query encoding process to handle number distributions. The detailed equations of the implementations of these backbones are given in Appendix~\ref{apdx:qe_instantiations}. 

\begin{itemize}
    \item GQE~\cite{hamilton2018embedding}: the graph query encoding model encodes a complex query into a vector in embedding space;
    \item Q2B~\cite{ren2020query2box}: the graph query encoding model encodes a complex query into a hyper-rectangle embedding space.
    \item Q2P~\cite{bai2022query2particles}: the graph query encoding model encodes a complex query into an embedding space with multiple vectors.
\end{itemize}
To the best of our knowledge, there are no privacy protection methods in NGDBs. Therefore, we compare our methods with noise disturbance which is similar to differential privacy~\cite{kasiviswanathan2013analyzing, qin2017generating}, a commonly employed technique in database queries that introduces randomness to answers through the addition of noise.

\subsubsection{Evaluation Metrics}
The evaluation consists of two distinct parts: reasoning performance evaluation and privacy protection evaluation. Following the previous work~\cite{bai2023knowledge},  given a testing query $q$, the training, validation, and public testing answers are denoted as $M_{train}$, $M_{val}$, and $M_{test}$, respectively. We evaluate the quality of retrieved answers using Hit ratio (HR) and Mean reciprocal rank (MRR). HR@K metric evaluates the accuracy of retrieval by measuring the percentage of correct hits among the top K retrieved items. The MRR metric evaluates the performance of a ranking model by computing the average reciprocal rank of the first relevant item in a ranked list of results. We evaluate the generalization capability of models by calculating the rankings of answers that cannot be directly retrieved from an observed knowledge graph, which is $M_{test}/ M_{val}$.
For reasoning performance evaluation, higher metric values denote better retrieval quality. For privacy protection evaluation, we compute the metric of privacy-threatening answers as these answers cannot be inferred from the observed graphs. Because we want to obfuscate those answers, lower values denote stronger protection. We train all the models by using the training queries and private attributes and tune hyper-parameters using the validation queries. The evaluation is then finally conducted on the testing queries, including the evaluation of public answers for performance assessment and the evaluation of private answers for privacy assessment. The experiment results are reported on the testing public and private queries respectively. 

\begin{table}[t]
\footnotesize
\caption{The main experiment results of public answers retrieval and private answers protection. NGDB can provide effective protection while sacrificing acceptable performance. }
\label{tab:general_performance}
\begin{tabular}{c|c|c|rr|rr}
\toprule
\multirow{2}{*}{Dataset} & \multirow{2}{*}{Encoding} & \multicolumn{1}{c|}{\multirow{2}{*}{Model}}  & \multicolumn{2}{c|}{Public} & \multicolumn{2}{c}{Private} \\ 
 & & & HR@3 & MRR & HR@3 & MRR \\
 \cmidrule{1-7}
\multirow{9}{*}{FB15k-N} & \multirow{3}{*}{GQE} & Baseline & \textbf{21.99} & \textbf{20.26} & 28.99 & 27.82 \\ 
 &  & Noise & 15.89 & 14.67 & 21.54 & 21.37 \\
  &  & P-NGDB & 15.92 & 14.73 & \textbf{10.77} & \textbf{10.21} \\   \cmidrule{2-7} 
 & \multirow{3}{*}{Q2B} & Baseline & \textbf{18.70} & \textbf{16.88} & 30.28 & 28.98 \\ 
 &  & Noise & 12.34&12.19&20.01&19.71  \\
  &  & P-NGDB & 12.28 & 11.18 & \textbf{10.17} & \textbf{9.38}\\   \cmidrule{2-7} 
 & \multirow{3}{*}{Q2P} & Baseline & \textbf{26.45} & \textbf{24.48} & 29.08 & 31.85 \\ 
 &  & Noise & 20.13 & 19.77 & 22.35 & 23.17  \\
  &  & P-NGDB & 19.48 & 18.19 & \textbf{14.15} & \textbf{14.93}\\  \cmidrule{1-7} 
\multirow{9}{*}{DB15k-N} & \multirow{3}{*}{GQE} & Baseline & \textbf{24.16} & \textbf{22.37} & 39.26 & 37.25 \\ 
 &  & Noise & 18.01 &16.35 & 28.59 & 28.37  \\
  &  & P-NGDB & 17.58 & 16.29 & \textbf{10.52}& \textbf{10.79}\\ 
   \cmidrule{2-7}
   & \multirow{3}{*}{Q2B} & Baseline & \textbf{15.94} & \textbf{14.98} & 42.19 & 39.78 \\ 
 &   & Noise & 10.76 & 10.28 & 26.49 & 25.93 \\
  &  & P-NGDB & 10.19 & 9.49 & \textbf{8.92 }& \textbf{7.99} \\   \cmidrule{2-7} 
 & \multirow{3}{*}{Q2P} & Baseline & \textbf{25.72} & \textbf{24.12} & 46.18 & 43.48 \\ 
 &  & Noise & 19.89 & 19.32 & 33.56 & 33.17  \\
  &  & P-NGDB & 20.26 & 19.00 & \textbf{19.38} & \textbf{18.45}\\ \cmidrule{1-7}
\multirow{9}{*}{YAGO15k-N} & \multirow{3}{*}{GQE} & Baseline & \textbf{26.06} &\textbf{24.37} & 43.55 & 40.81 \\ 
 &  & Noise & 20.32 & 20.27 & 38.52 & 38.29  \\
  &  & P-NGDB & 19.58 & 19.82 & \textbf{7.56} & \textbf{7.33}\\ \cmidrule{2-7}
  
  & \multirow{3}{*}{Q2B} & Baseline & \textbf{23.39} & \textbf{22.53} & 42.73 & 40.55 \\ 
 &  & Noise & 16.85 & 15.37 & 28.23 & 28.54  \\
  &  & P-NGDB & 17.07 & 16.03 & \textbf{6.26} & \textbf{5.79}\\   \cmidrule{2-7} 
 & \multirow{3}{*}{Q2P} & Baseline & \textbf{29.41} & \textbf{27.87 }& 42.56 & 45.79\\ 
 &  & Noise & 22.85 & 21.21 & 34.26 & 33.68 \\
  &  & P-NGDB & 23.27 & 22.59 &\textbf{7.34}& \textbf{7.17}\\ 
\bottomrule
\end{tabular}
\vspace{-0.4cm}
\end{table}

\begin{table*}[t]

\caption{The mean reciprocal ranking (MRR) variation for P-NGDBs on different types of queries. The performance ratio to the corresponding unprotected models is reported in parentheses.}
\vspace{-0.3cm}
\small
\label{tab:detailed_performance}
\begin{tabular}{cc|l|rrrrrrrr}
\toprule
\multicolumn{1}{c|}{Dataset} & Enc. & Test.  & 1p & 2p & 2i & 3i & pi & ip & 2u & up \\ \midrule
\multicolumn{1}{c|}{\multirow{6}{*}{FB15k-N}} & \multirow{2}{*}{GQE} & Public & 10.51(70.4\%) & 3.89(55.3\%) & 23.49(89.1\%) & 44.14(92.2\%) & 14.57(83.4\%) & 6.43(73.5\%) & 4.15(59.2\%) & 2.95(70.9\%) \\  
 \multicolumn{1}{c|}{} &  & Private& 0.18(4.6\%) & 1.80(13.6\%) & 11.77(45.7\%) & 32.71(56.2\%) & 14.07(46.6\%) & 6.37(36.7\%) & 0.70(6.1\%) & 1.03(13.6\%) \\
  \cmidrule{2-11} 
\multicolumn{1}{c|}{} & \multirow{2}{*}{Q2B} & Public & 12.88(69.8\%) & 3.27(53.6\%)& 20.13(85.3\%) & 36.52(83.4\%) & 12.18(69.7\%) & 4.57(59.1\%) & 5.41(57.8\%) & 3.04(62.5\%)\\  
\multicolumn{1}{c|}{}  &  & Private & 0.41(11.6\%) & 3.02(23.4\%) & 9.45(31.5\%) & 27.42(50.3\%) & 15.21(47.5\%) & 9.49(47.1\%) & 0.23(5.0\%) & 1.48(19.1\%) \\
  \cmidrule{2-11} 
\multicolumn{1}{c|}{} & \multirow{2}{*}{Q2P} & Public & 16.36(73.5\%) & 6.22(59.8\%) & 23.34(82.5\%) & 43.29(93.6\%) & 15.26(77.3\%) & 7.44(63.8\%) & 6.79(60.8\%) & 3.64(58.2\%) \\  
 \multicolumn{1}{c|}{} &  & Private & 0.31(11.3\%) & 3.24(31.3\%) & 15.82(46.7\%) & 39.49(46.7\%) & 18.12(52.7\%) & 9.60(45.1\%) & 0.38(8.1\%) & 2.55(23.3\%)\\ 
\hline
\multicolumn{1}{c|}{\multirow{6}{*}{DB15k-N}} & \multirow{2}{*}{GQE} & Public & 2.35(58.1\%) & 0.98(56.8\%)  & 22.30(83.1\%)  & 49.54(85.4\%)  & 10.06(67.0\%)  & 3.48(53.0\%)  & 0.73(51.2\%)  & 0.94(57.2\%)  \\  
\multicolumn{1}{c|}{} &  & Private & 0.22(9.3\%) & 1.76(17.4\%)  & 12.33(17.7\%)  & 31.88(44.9\%)  & 23.26(57.9\%)  & 4.83(47.9\%)  &0.76(12.1\%)  & 0.94(19.2\%)  \\
   \cmidrule{2-11} 
  \multicolumn{1}{c|}{} & \multirow{2}{*}{Q2B} & Public & 2.38(53.0\%)  & 1.54(65.8\%)  & 16.16(83.5\%)  & 28.44(85.1\%)  & 9.23(63.9\%)  & 2.68(52.3\%)  & 0.81(57.2\%)  & 0.98(55.4\%)  \\
\multicolumn{1}{c|}{}  &  & Private& 0.18(7.6\%)  & 2.72(26.8\%)  & 9.34(13.4\%)  & 24.89(34.7\%)  & 19.10(53.3\%)  & 6.06(53.0\%)  & 0.74(12.5\%)  & 0.67(16.0\%)  \\
  \cmidrule{2-11} 
\multicolumn{1}{c|}{} & \multirow{2}{*}{Q2P} & Public & 4.75(60.3\%)  & 2.59(61.8\%)  & 24.67(80.2\%)  & 52.48(94.8\%)  & 13.41(70.2\%)  & 4.62(48.1\%)  & 2.20(54.3\%)  & 1.58(46.8\%)  \\  
 \multicolumn{1}{c|}{} &  & Private & 0.23(8.7\%)  & 2.92(28.4\%)  & 28.37(39.5\%)  & 35.58(47.0\%)  & 24.55(57.4\%)  & 9.05(57.6\%)  & 1.62(23.9\%)  & 1.87(22.5\%)  \\
\midrule
\multicolumn{1}{c|}{\multirow{6}{*}{YAGO15k-N}} & \multirow{2}{*}{GQE} & Public& 5.28(86.5\%) & 1.78(57.9\%) & 24.35(66.9\%) & 43.2(65.0\%) & 13.98(70.3\%) & 4.32(60.7\%) & 3.56(89.2\%) & 1.03(66.0\%) \\  
 \multicolumn{1}{c|}{} &  & Private & 0.12(6.3\%) & 3.21(15.0\%) & 12.56(37.2\%) & 27.65(35.6\%) & 14.09(45.2\%) & 12.87(53.7\%) & 1.08(7.9\%) & 1.20(9.3\%) \\ 
 \cmidrule{2-11} 
\multicolumn{1}{c|}{} & \multirow{2}{*}{Q2B} & Public & 6.06(74.3\%)  & 1.98(51.5\%)  & 24.03(72.2\%)  & 44.52(81.7\%)  & 13.22(63.3\%)  & 4.33(52.4\%)  & 3.84(61.9\%)  & 1.11(56.1\%) \\  
 \multicolumn{1}{c|}{} &  & Private & 0.11(5.6\%) & 3.45(16.6\%) & 16.99(50.4\%) & 20.62(32.4\%) & 13.51(39.1\%) & 14.42(48.3\%) & 1.32(10.1\%) & 1.14(12.3\%)\\ 
  \cmidrule{2-11} 
\multicolumn{1}{c|}{} & \multirow{2}{*}{Q2P} & Public & 9.96(83.7\%) & 3.78(67.7\%) & 30.96(77.1\%) & 37.87(68.9\%) & 15.23(64.3\%) & 6.59(64.9\%) & 9.63(71.4\%) & 2.18(66.1\%) \\  
 \multicolumn{1}{c|}{} &  & Private & 0.12(9.0\%) & 3.17(18.1\%) & 14.29(35.0\%) & 34.29(42.6\%) & 16.59(45.2\%) & 13.95(47.2\%) & 1.96(9.9\%) & 2.68(13.6\%) \\
  \midrule
  \multicolumn{2}{c|}{\multirow{2}{*}{Avg. Change}} & Public & $\downarrow$ 30.0\%& $\downarrow$ 41.1\%  & $\downarrow$ 20.0\% & $\downarrow$ 16.7\% & $\downarrow$ 30.1\%& $\downarrow$ 41.4\% & $\downarrow$ 37.4\%& $\downarrow$ 40.1\%\\
   & & Private&$\downarrow$ 91.8\% & $\downarrow$ 78.8\%& $\downarrow$ 59.9\%& $\downarrow$ 56.7\%& $\downarrow$ 50.6\%& $\downarrow$ 51.5\%& $\downarrow$ 67.1\%& $\downarrow$ 83.5\% \\
 
\bottomrule
\end{tabular}
\vspace{-0.3cm}
\end{table*}

\subsubsection{Parameter Setting.} We first tune hyper-parameters on the validation queries for the base query encoding methods. We use the same parameters for NGDBs for a fair comparison. We tune the privacy penalty coefficient $\beta$ for three datasets respectively for utility privacy tradeoff for convenient illustration. For noise disturbance privacy protection, we adjust the noise strength to make the results comparable to P-NGDBs' results (similar performance on public query answers).

\subsection{Performance Evaluation}

We apply our P-NGDB on three different query encoding methods GQE~\cite{hamilton2018embedding}, Q2B~\cite{ren2020query2box}, and Q2P~\cite{bai2022query2particles} and compare the query answering performance with and without P-NGDB's protection. Experiment results are summarized in Table \ref{tab:general_performance} and Table \ref{tab:detailed_performance}. Table \ref{tab:general_performance} reports the averaged results in HR@3 and MRR. A higher value on public answer sets denotes better reasoning ability, while a smaller value on private answer sets denotes stronger protection. The experiment results show that without protection, the base encoding methods all suffer from privacy leakage problems if there is sensitive information existing in the knowledge graph. Our proposed P-NGDB can effectively protect private information with a slight loss of complex query answering ability. For example, in FB15K-N, GQE retrieves the public answers with 21.99 HR@3 and the private answers with 28.99 HR@3 without privacy protection. With P-NGDB's protection, GQE retrieves public answers with 15.92 HR@3 and private answers with 10.77 HR@3. The private answers are protected with a 62.9\% decrease in HR@3 sacrificing a 27.4\% decrease in HR@3 for public answers. Compared to the noise disturbance protection, our method can accurately protect sensitive information, so that the loss of performance is much lower than the noise disturbance. 

In Table \ref{tab:detailed_performance}, we show P-NGDBs' MRR results on various query types and the percentages within parentheses represent the performance of the P-NGDB protected model relative to the performance of the encoding model without protection. The evaluation is conducted on public and private answers respectively to assess privacy protection. From the comparison between MRR variation on public and private answers, we can know that P-NGDB can provide protection on all the query types. Besides, comparing the average change on these query types, we can know from the results that the P-NGDB have different protection ability on various query types. For projection and union, the MRR of protected models on private answers is largely degraded, denoting a stronger protection. While for the intersection operator, the preservation provided by P-NGDBs is as effective as protections for other operators.
\begin{figure}[t]
  \centering
  \includegraphics[width=0.9\linewidth]{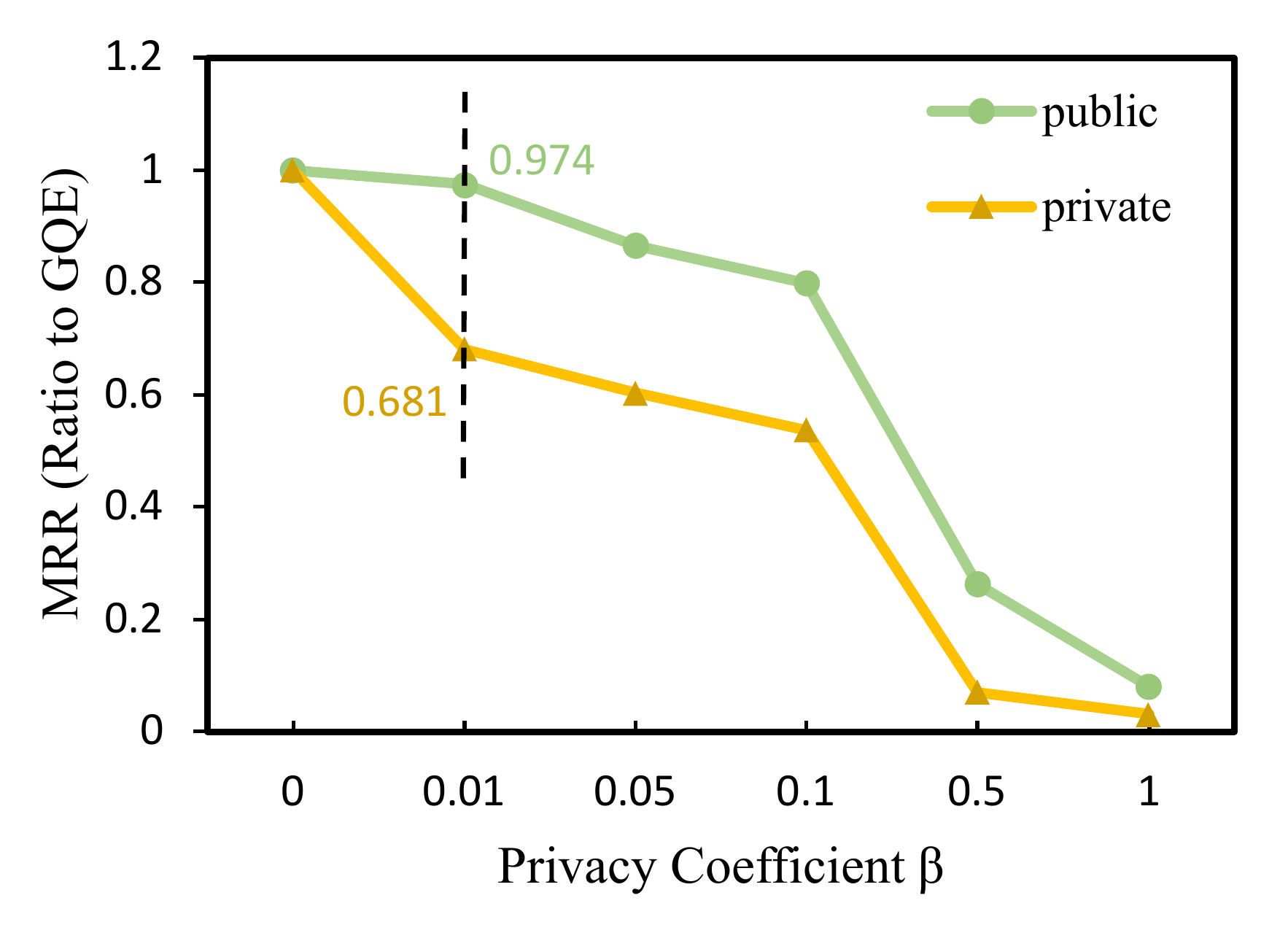}
    \vspace{-0.3cm}
  \caption{The evaluation results of GQE with various privacy coefficients $\beta$ on FB15K-N.}
  \label{fig:penalty}
  \vspace{-0.4cm}
\end{figure}
\subsection{Sensitivity Study}
The privacy protection needs can be different in various knowledge graphs. We evaluate the impact of the privacy penalty coefficient $\beta$,  A larger $\beta$ can provide stronger privacy protection. We select $\beta$ from $\{0.01, 0.05, 0.1, 0.5, 1\}$, and evaluate the retrieval accuracy of public answers and private answers, respectively. We evaluate GQE's retrieval performance on the FB15K-N dataset and depict the MRR's change under each $\beta$ compared to the unprotected GQE model. The results are shown in Figure~\ref{fig:penalty}. We know that the penalty can effectively control the level of privacy protection. For example, when $\beta=0.01$, the GQE model can achieve 97.4\% MRR on public answers while only having 68.1\% MRR on private answers. The flexible adjustment of the privacy coefficient can make P-NGDBs adapt to various scenarios with different privacy requirements. However, to meet higher privacy needs, P-NGDBs have to undertake much utility loss.

%% file: conclusion.tex
\section{Conclusion \label{sec:conclude}}

In this work, we proposed the privacy problem in neural graph databases and showed that sensitive information can be inferred by attackers with specified queries. We denote that some query answers can leak privacy information named privacy risk answer. To systematically evaluate the problem, we constructed a benchmark dataset based on FB15k-N, YAGO15k-N, and DB15k-N.  Finally, we proposed a new framework that can protect the privacy of the knowledge graph from attackers' malicious queries.  Experimental results on the benchmark shows that the proposed model (P-NGDB) can effectively protect privacy while sacrificing slight reasoning performance. In the future, we will take logical operators into consideration and make improvements in NGDB privacy protection.

%% file: appendix.tex
\appendix

\section{Instantiations of Query Encoders}
In this section, we present detailed formulas for parameterizing various logical operations utilized in this paper. Additionally, we employ the following three query encoders: GQE, Q2B, and Q2P, as the foundational components of our approaches.
\label{apdx:qe_instantiations}
\subsection{Graph Query Encoding (GQE)}
In the GQE method~\cite{hamilton2018embedding}, every query is represented as a singular vector. We denote the query embedding of each sub-query as $q_i \in \mathbf{R}^k$. In logical operations that involve multiple inputs, if there are multiple sub-queries, we denote them as $q_i^1, q_i^2, ..., q_i^n$. Consistent with the paper, we employ the TransE variant of GQE for the instantiation of the relational projection in GQE. The specifics are outlined below.
The equation for updating the query embedding in the GQE method is given by:
\begin{equation}
q_{i+1} = f_P(q_i, r) = q_i + e_r,
\end{equation}
where $e_r$ represents the relation embedding of $r$.

Additionally, for modeling the intersection operation, we utilize feed-forward functions and average pooling to implement a permutation-invariant neural network.
\begin{equation}
q_{i+1} = f_I(q_i^1,...,q_i^n) = W_I(\phi( \texttt{FFN}(q_i^k), \forall k \in \{1,2,...,n \} )),
\end{equation}
where \texttt{FFN} is a feed-forward function, $\phi$ is an average pooling, and $W_I$ is parameter matrix. Following the setting in Q2B paper~\cite{ren2020query2box}, we express the queries in disjunction normal form so that we can avoid the parameterization of union operations. 
\subsection{Box Embeddings (Q2B)}
For the Q2B model, we implement the operations following the original paper. The box embedding can be separated into two parts, denoting the center and offsets, namely $q_i = [c_i, o_i]$. Meanwhile, for each relation $r$, there is a corresponding box relation embedding $e_r = [c_r, o_r]$. 
The relation projection $q_{i+1} = f_P(q_i, r)$ is computed as follows
\begin{equation}
q_{i+1} = q_{i} + e_r = [c_i+c_r, o_i + o_r]
\end{equation}
The intersection operations $q_{i+1} = f_I( q_i^1, q_i^1, ..., q_i^n )$ is computed as follows 
\begin{align}
    c_{i+1} & = \sum_k a_k \cdot c_i^k, \thickspace a_k = \frac{\texttt{exp}(\texttt{MLP}(q_k))}{\sum_j \texttt{exp}(\texttt{MLP}(q_j))} \\
    o_{i+1} & = \texttt{Min}(\{ o_i^1, o_i^2, ..., o_i^n \}) \cdot \sigma (\texttt{DeepSets}(\{q_i^1, q_i^2, ..., q_i^n \})),
\end{align}
then the query embedding of $i+1$ step is $q_{i+1} = [c_{i+1}, p_{i+1}]$.
In this equation, $\cdot$ is the dimension-wise product, \texttt{MLP} is a multi-layer perceptron, and $\sigma$ is the sigmoid function. $\texttt{DeepSets}(\cdot)$ is the permutation-invariant deep architecture, which treats all the input sub-queries equally. The $\texttt{DeepSets}(\{x_1, x_2,..., x_N\})$ is implemented as $\texttt{MLP}(\frac{1}{N}\sum_i(\texttt{MLP}(x_i)))$.
Following the original paper~\cite{ren2020query2box}, we use the disjunction normal form to avoid the direct parameterization of the union operation.

\subsection{Particle Embeddings (Q2P)}
In the Q2P embedding method~\cite{bai2022query2particles}, query embeddings are represented as a set of particle embeddings. Specifically, the query embedding can be represented as $q_i = [p_i^1, p_i^2, ..., p_i^k]$, where $k$ denotes the number of particles in the embedding.

The relation projection $f_P$ is defined as follows:
$q_{i+1} = f_P(q_i, e_l)$,
where $q_i$ and $q_{i+1}$ represent the input and output particle embeddings, respectively.

Unlike the approach in \cite{bordes2013translating}, where the same relation embedding $e_l$ is directly added to all particles in $q_i$ to model the relation projection, Q2P introduces multiple neuralized gates. These gates enable individual adjustment of the relation transition for each particle in $q_i$. The formulation is as follows:
\begin{align}
&Z = \sigma(W^{P}_ze_l + U_zq_i  + b_z), \\
&R = \sigma(W^{P}_re_l + U_rq_i + b_r), \\
&T = \phi(W^{P}_he_l + U_h(R \odot q_i ) + b_h),\\
&A_i = (1-Z) \odot q_i + Z \odot T.
\end{align}
Here,  $\sigma$ and $\phi$ are the sigmoid and hyperbolic tangent functions, and  $\odot$ is the Hadamard product.
To allow information exchange among different particles, a scaled dot-product self-attention \citep{vaswani2017attention} module is also incorporated,
\begin{align}
q_{i+1} = \texttt{Attn}(W^{P}_q A^T_i , W^{P}_k A^T_i, W^{P}_v A^T_i )^T.
\end{align}
The parameters $W^{P}_q, W^{P}_k, W^{P}_v \in \mathbb{R}^{d \times d}$ are utilized to model the input Query, Key, and Value for the self-attention module called \texttt{Attn}.
The intersection operation $f_I$ is defined on multiple sets of particle embeddings $\{q^{(n)}_i\}_{n=1}^N$.
It outputs a single set of particle embeddings $q_{i+1} = f_I(\{q^{(n)}_i \}_{n=1}^N)$.
The particles from the $\{q^{(n)}_i \}_{n=1}^N$ are first merged into a new matrix $q_i = [q^{(1)}_i, q^{(2)}_i, ...,q^{(N)}_i]\in R^{d \times NK}$, 
and this matrix $q_i$ serves as the input of the intersection operation. 
The operation updates the position of each particle based on the positions of other input particles. This process is modeled using scaled dot-product self-attention, followed by a multi-layer perceptron (MLP) layer. The equations for this process are as follows:

\begin{align}
& A_i = \texttt{Attn}(W^{I}_{q}q^T_i, W^{I}_kq^T_i, W^{I}_vq^T_i)^T, \\
& q_{i+1} = \texttt{MLP}(A_i).
\end{align}

Here, $W^{I}_q, W^{I}_k, W^{I}_v \in \mathbb{R}^{d \times d}$ are the parameters for the self-attention layer. The \texttt{MLP} represents a multi-layer perceptron layer with ReLU activation. It is important to note that the parameters in the \texttt{MLP} layers of different operations are not shared.